\documentclass[twocolumn,prl,superscriptaddress,floatfix]{revtex4}
\usepackage{graphicx}
\usepackage{xcolor}
\usepackage{epsfig}
\usepackage{rotating}
\usepackage{amsmath}
\usepackage{amsfonts}
\usepackage{amssymb}
\usepackage{enumerate}
\usepackage{longtable}
\usepackage{appendix}
\usepackage{dcolumn}
\usepackage{bm}

\usepackage{hyperref}
\usepackage{ulem}

\usepackage{braket}

\begin{document}

\title{Magnetism of competing high-spin/low-spin states in Ba$_2$NiO$_2$(AgSe)$_2$ and related two-orbital two-electron systems}

\author{Rajiv R. P. Singh}
\affiliation{Department of Physics, University of California Davis, 
CA 95616, USA}

\date{\rm\today}

\begin{abstract}
 We discuss the thermodynamic and magnetic properties of a competing high-spin/low-spin two-orbital two-electron model on a square-lattice possibly relevant to the nickelates such as Ba$_2$NiO$_2$(AgSe)$_2$ (BNOAS). We focus on parameter regimes where a high-spin ($S=1$) and a low-spin ($S=0$) state are energetically close to each other and discuss various exchange processes in such a system. The model we study is a variant of, but different from the Kondo-necklace model proposed for the system by Jin et al \cite{jin}. Although there are similarities between the two models in terms of the ground-state phases and their symmetries, the detailed properties of the different phases and the phase transitions are entirely different and should be easy to distinguish from an experimental point of view.
\end{abstract}


\maketitle

\section{I. Introduction}

Following the discovery of high-temperature superconductivity in the cuprates, many new families of materials have been synthesized with competing magnetic and superconducting phases. In particular, several iron based superconductors are known to have competing magnetic and superconducting phases and, at stoichiometry, are often modelled as multi-orbital 
spin-one systems \cite{qimiao}. 

The discovery of superconductivity in the nickelates has brought renewed interest to these systems as well \cite{hwang}. Although much recent excitement comes from doping the Nd$_{n+1}$Ni$_n$O$_{2n+2}$ family of materials where nickel ions have a Ni$^{1+}$ ionic state with $d^9$ electron configuration providing an analog of the cuprates, there are other nickelate families with the more usual $Ni^{2+}$ ionic state and $d^8$ electron configuration \cite{klee2,botana}. Our work is focused on competing high-spin and low-spin states in these latter systems. 

Jin et al \cite{jin,klee} have recently studied the material Ba$_2$NiO$_2$(AgSe)$_2$ (BNOAS) by Density Functional Theory (DFT) and found a rather striking result. Ni atoms have $d^8$ electronic configuration, with competing spin-zero and spin-one states, where the lowest energy state has spin-zero. Depending on the value of the Hubbard repulsion $U$, this lowest energy singlet state can be made of two electrons occupying two different orbitals ($d_{x^2-y^2}$ and $d_{z^2}$). A lower energy of this singlet compared with the triplet amounts to an {\it antiferromagnetic} intra-atomic exchange, which goes against the Hund's rules. Jin et al have proposed an effective Kondo-necklace model, where the two electrons on an atom are exchange coupled by an antiferromagnetic intra-atomic exchange. In addition, one of the electrons (in the $d_{x^2-y^2}$ orbital) has an exchange interaction with electrons on neighboring atoms in the plane, whereas the electron in the other state (in the $d_z^2$ orbital) provides an analog of the local spin in the Kondo models. 
To leading order, its only coupling is to the other spin in the same atom. This leads to a Kondo-necklace model with two spin-half operators $\vec S_{1i}$ and $\vec S_{2i}$ at each site of the square-lattice and Hamiltonian:
\begin{equation}
H = J \sum_{\langle i,j \rangle} \vec S_{1i} \cdot \vec S_{1j}
+ J_{l} \sum_{i} \vec S_{1i} \cdot \vec S_{2i},
\label{K-Hamiltonian}
\end{equation} 
where the first sum runs over nearest-neighbor pairs.
At small $J/J_{l}$ the system is in a non-magnetic singlet phase, where the two spins on each atom combine to form a singlet. At large $J/J_{l}$, the system is in the Neel phase with antiferromagnetic order. The two phases are separated by a quantum critical point in the universality class of the 3-dimensional classical Heisenberg model \cite{brenig}.

Here we will focus on an alternative model for this insulating system with $d^8$ electron configuration and two d-holes in two orbitals with competing high-spin and low-spin states. In a two-orbital system, two electron (or two-hole) configuration leads to one triplet and 3 singlets. Two of the singlets correspond to the two electrons occupying the same orbital, while a singlet and a triplet correspond to electrons occupying different orbitals. We will assume that the intra-atomic Hund's coupling is the normal {\it ferromagnetic} one so that the triplet state is lower in energy when the two electrons occupy different orbitals. However,  Hubbard repulsion $U$, crystal field splittings and ligand interactions are such that one of the singlets formed from holes occupying the same orbital (plus ligands) is lower in energy than  the triplet state. Since intra-atomic Hund's exchange is typically of order $1$ eV, we will assume that the other two singlet states are significantly higher in energy and not relevant to the low energy properties. A similar competing low-spin/high-spin scenario has been discussed recently for Ni $d^8$ ions in the material NiO$_2$ by Jiang et al \cite{sawatzky}.

\section{II. Exchange processes and the Model Hamiltonian} 
The overlap between orbitals on neighboring atoms leads to inter-atomic exchange processes.  A general two-band Hubbard model for the nickelates was studied by Hu and Wu \cite{wu}, who used perturbation theory to argue for a Kugel-Khomskii type spin-orbital model \cite{khomskii} with many possible terms.

Here we take as our starting point for deriving the effective low energy Hamiltonian two key assumptions motivated by the work of Jin et al \cite{jin} and Jiang et al \cite{sawatzky}: (i) One spin-zero and one spin-one state forms the low energy subspace of our system and all others states are at significantly higher energy. Thus, all inter-atomic exchanges must be projected on to these low energy states of each atom. (ii) Inter-atomic hopping is much larger for spins in one of the orbitals (e.g. $d_{x^2-y^2}$) than the other (e.g. $d_{z^2}$). These considerations lead to the following inter-atomic exchanges in order of decreasing magnitude:

\textit{1. Spin-one Heisenberg exchange J}: The largest inter-atomic exchange is a Heisenberg coupling J between neighboring atoms when both are occupied by spin-one states. For this exchange, overlap between $d_{x^2-y^2}$ orbitals is sufficient. The difference from the Kondo necklace model of Jin et al is that the projection to the low energy subspace means that the Heisenberg coupling is between spin-one states on neighboring atoms. It is not a coupling for spin-half or just one of the two spins. That is forbidden by the large intra-site Hund's coupling.

\textit{2. Triplet hopping process K}: When there is a spin-zero on one site and a spin-one on a neighboring site, the two sites can exchange their spin states. This is a much weaker process than the Heisenberg exchange as it requires overlap of both (e.g. $d_{x^2-y^2}$ and $d_{z^2}$ ) orbitals on neighboring atoms and thus may require either a lattice distortion or a higher-order process. In our model we will ignore this term in the numerical study. Qualitatively, it is important in the non-magnetic phase where even a small value is relevant to the dispersion of spin excitations.

\textit{3. Triplet-pair creation/destruction process W}: A pair of triplets with total spin-zero can turn into a pair of spin-zero states on neighboring sites and vice versa. This process will require cross overlap between neighboring $d_{x^2-y^2}$ and $d_{z^2}$ orbitals. We will assume this process is the weakest and will not discuss it any further. 

Thus, our low energy states on an atom are characterized by $n_i=0$ or $1$, with $n_i=0$ corresponding to spin-zero and $n_i=1$ to the spin-one state. In addition, when $n_i=1$, there are three spin states corresponding to $S^z_i=0, \pm 1$. Thus, there are 4-states per atom. The effective Hamiltonian can be expressed in terms of $n_i$ and the quantum spin-one operators $\vec S_i$ as:
\begin{equation}
H = \Delta \sum_i n_i + J \sum_{\langle i,j \rangle} n_i n_j \vec S_{i} \cdot \vec S_{j},
\label{Hamiltonian}.
\end{equation} 
We will assume that the parameter $\Delta$ is greater than zero so that the lowest energy state on each atom has spin-zero.

\section{III. Ground State Phases of the model}
Ground state phases and properties of the Kondo-necklace model have been studied previously
\cite{brenig}. Certain aspects of the ground state phase diagram of the present model can be obtained using results from the literature \cite{singh,hamer,hamer2}.
Like the Kondo-necklace model, this model has two phases. For small $J/\Delta$ the model is in a non-magnetic singlet phase, where each atom is occupied by electrons in a spin-zero state. For large $J/\Delta$ the model is in a {\it spin-one} Neel state with long-range antiferromagnetic order. However, there are key differences in the properties of the two phases and the phase-transition between the two models.

\textit{Neel phase}: The Neel phase is characterized by long-range antiferromagnetic order and magnon excitations. The key difference between the models lies in the behavior of the uniform susceptibility. In the present model, deep in the Neel phase, the susceptibility has the temperature dependence of an antiferromagnet, that goes down below a temperature of order $J$, when short-range antiferromagnetism develops. At low temperatures, as the spins are locked antiferromagnetically, they have only a weak transverse susceptibility \cite{singh,hamer,hamer2}. In contrast, in the Kondo-necklace model, as one gets away from the critical point, the low temperature uniform susceptibility becomes large. This is because the local spins become nearly free and thus develop almost Curie-like susceptibility \cite{brenig}.

\textit{Non-magnetic singlet phase}: In the Kondo necklace model, the singlet phase has a dispersive triplet excitation, where the dispersion band-width is set by the inter-atomic exchange energy $J$. As the phase transition is approached, the excitation gap will go to zero at the antiferromagnetic wave-vector and at $k=0$, with neutron-scattering spectral weight strongly concentrated near the anti-ferromagnetic wave-vector. In contrast, for the present model, the triplet will be static and completely localized in the absence of the triplet-hopping process discussed earlier. It is the weaker triplet hopping process $K$ that will lead to a dispersion for the excitations. Thus, one would expect the dispersion to be relatively weak and being unrelated to the exchange $J$ it should have a minimum at $k=0$. In addition, there will be a clustering of the spinful states as when the spin-one states are clustered together their energy will be lowered than when they are separated.

\textit{Ground state Phase Transition}: In the Kondo necklace model there is a quantum critical point separating the non-magnetic phase from the antiferromagnet. In contrast, in the present model, there will be a first order phase transition. Well before the gap to spin-one excitations closes, the exchange energy will lower the energy of the Neel state below that of singlets. Thus the system will undergo a phase separation and a first order ground-state phase-transition between the two phases. Since the ground state energy of the spin-one model is known accurately \cite{singh} to be approximately $-2.327 J$ per site, the phase transition will happen at $\Delta=2.327 J$.

\begin{figure}[htb!]
\centering
\includegraphics[width=\columnwidth]{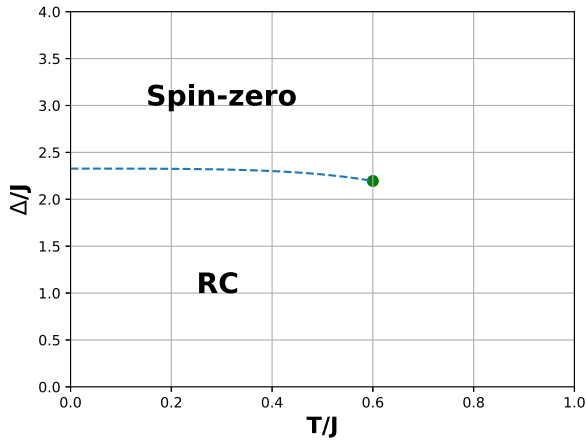}
\caption{Sketch of the phase diagram for the model. A line of first order phase transitions, shown by dashed line, separates the Renormalized Classical phase (RC) from the non-magnetic spin-zero phase. The line of phase transitions terminates at an Ising critical point.}
\label{phase-d} 
\end{figure}

\textit{Finite temperature phase transition}: In the present model there will be a first order liquid-gas type phase transition at finite temperatures ending at a critical point. Even though the spin degrees of freedom cannot order at finite temperatures, the Ising variables $n_i$ can have a true phase transition. Since the transition will occur at temperatures well below the mean-field Neel  temperature of $8 J/3$, by that temperature short-range order will be well developed. Hence, we can study the transition to a good approximation by replacing the Heisenberg coupling in the Hamiltonian by its low temperature expectation value which will be only weakly temperature dependent. Thus our model becomes:
\begin{equation}
H = \Delta \sum_i n_i -{\Tilde J} \sum_{\langle i,j \rangle} n_i n_j,
\label{Liquid-gas}
\end{equation} 
where ${\Tilde J}$ is minus the expectation value of $<\vec S_i \cdot \vec S_j>$ on a nearest neighbor bond. It is weakly temperature dependent at low temperatures \cite{singh}
\begin{equation}
 {\Tilde J}(T) ={2.327\over 2} J - a T^3.
\label{JT}
\end{equation} 
The $T^3$ term comes from the reduction in energy due to the magnons.

By the standard mapping
\begin{equation}
    n_i= (s_i +1)/2,
\end{equation}
this can be expressed in terms of the Ising variables $s_i=\pm 1$ as:
\begin{equation}
H = - J_I \sum_{\langle i,j \rangle} s_i s_j
+ h \sum_i s_i + C,
\label{Ising}.
\end{equation}
where 
$$J_I={1\over 4} {\tilde J},$$ and, $$h={\Delta- 2{\tilde J}\over 2}.$$

For the Ising model in Eq.~\ref{Ising}, the phase transition line is given by $h=0$, that is $\Delta = 2 {\Tilde J}$. At $T=0$, the transition is at $\Delta=2.327 J$. The line of first order transitions will terminate at a critical point, which can be obtained from the solution of the 2D Ising model to be at $T= {2 J_{I}\over \ln{1+\sqrt{2}}}$. In the $\Delta/J$ and $T/J$ plane the phase boundary gradually bends down due to the temperature dependence of ${\Tilde J}$ and the critical point is roughly at $\Delta/J=2.2$, $T_c/J=0.6$. The phase diagram is sketched in Fig~1.

\section{IV. Numerical Study of finite temperature properties}

\begin{figure}[htb!]
\centering
\includegraphics[width=\columnwidth]{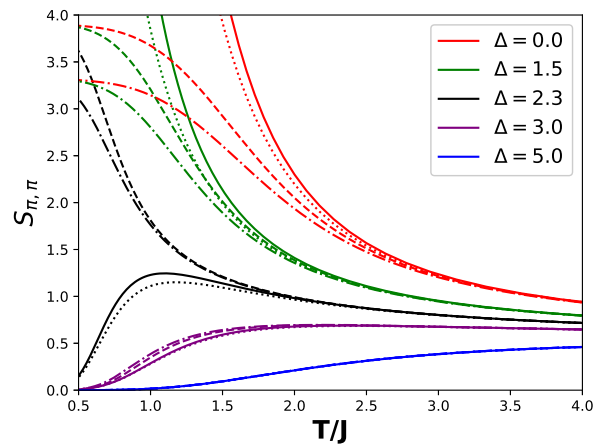}
\caption{Antiferromagnetic structure factors as a function of temperature for various values of $\Delta$. Solid lines are 6th order NLC, dotted lines are 5th order NLC, dashed lines are ED results for a 10-site cluster and dash-dotted lines are ED results for an 8-site cluster.}
\label{spi} 
\end{figure}

In this section we study properties of the model more quantitatively at high temperatures using Numerical Linked Cluster Expansions (NLC) \cite{rigol} and Exact Diagonalization (ED) of an 8 and a 10 site periodic cluster. In NLC, extensive properties on a lattice ${\cal L}$ with $N$ sites are calculated, in the thermodynamic limit $N\to\infty$, as a sum over all linked clusters $c$ as
\begin{equation}
    P({\cal L})/N =\sum_c L(c) \times W(c)
\end{equation}
Here $L(c)$, called the lattice constant, is the number of times the cluster arises in the lattice, per lattice site. The quantity $W(c)$ is the weight of the cluster. It is defined recursively through the property on the finite cluster $c$ as
\begin{equation}
    W(c) =P(c) - \sum_{s} W(s),
\end{equation}
where the sum runs over all proper sub-clusters $s$ of the cluster $c$. 

\begin{figure}[htb!]
\centering
\includegraphics[width=\columnwidth]{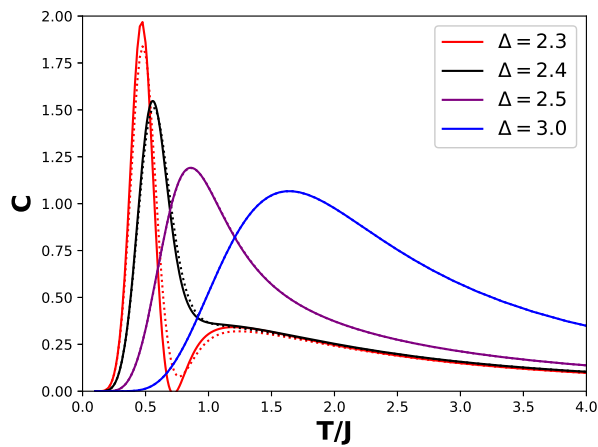}
\caption{Heat capacity as a function of temperature for different values of $\Delta$. Solid lines correspond to 6th order NLC and dotted lines to 5th order NLC.}
\label{c} 
\end{figure}

\begin{figure}[htb!]
\centering
\includegraphics[width=\columnwidth]{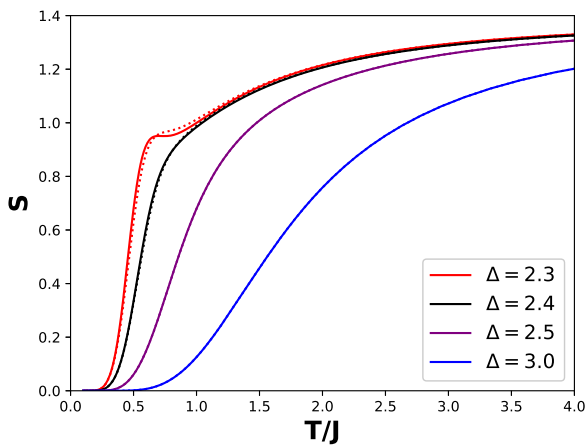}
\caption{Entropy as a function of temperature for different values of $\Delta$. Solid lines correspond to 6th order NLC and dotted lines to 5th order NLC.}
\label{s} 
\end{figure}

We calculate the heat capacity per site $C$, the entropy per site $S$, the uniform susceptibility per site $\chi$ and the antiferromagnetic structure factor $S_{\pi,\pi}$. Calculations are done to 6th order, that is weights of all linked clusters with 6 or less bonds are included in our study. NLC converges at high temperatures but will break down when correlation length exceeds a few lattice constants \cite{rigol}.

The parameter regimes of the model can be divided into 3 regions as seen from the behavior of the structure factor in Fig.~2. The first region is $-\infty<\Delta<2$. Representative plots for the structure factors are shown for $\Delta=0$ and $\Delta=1.5$. As $\Delta\to -\infty$ this model reduces to spin-one Heisenberg model which is known to have a renormalized classical (RC) phase at low temperatures. All the way up to $\Delta=2.0$ the fundamental behavior remains unchanged. The only change in properties is that at high temperature the existence of spin-zero states pushes down the onset of antiferromagnetic correlations. Note that NLC and ED agree above $T/J=3$. At lower temperature the finite size data must saturate, they are limited by total number of spins. NLC shows a rapid rise in the structure factor. At the lowest temperatures NLC must also break down as the correlation length increases exponentially.

The second region is $\Delta>2.5$. Representative plots for the structure factors are shown for $\Delta=3.0$ and $5.0$. This is a spin-zero phase in which NLC converges well at all temperatures. The only finite size effects in ED are at intermediate temperatures as seen at $\Delta=3.0$ and they are relatively small.

The third region is the transition region $2.0<\Delta<2.5$. This is a difficult region for numerical study at intermediate temperatures. As illustration, structure factor for $\Delta=2.3$ is shown. One can see that ED and NLC diverge below a temperature of $1.5 J$. Finite sizes are too small to capture the intermediate temperature behavior. Although NLC also breaks down as we get close to the transition, it does give us a glimpse of the transition behavior, which we now discuss. 

In Fig.~3 we show the heat capacity as a function of temperature for $\Delta=$ $2.3$, $2.4$, $2.5$ and $3.0$. As $\Delta$ is reduced from large values, the peak in the specific heat moves to lower temperature and it increases in magnitude. However, at $\Delta=2.3$ the specific heat curve develops a two-peak structure. The weak higher T peak signifies a development of spin-zero dominance over spin-one (gas phase in the liquid-gas analogy). This gives way, through a first order transition, to the antiferromagnetic phase (liquid or condensed phase in the liquid-gas analogy) at still lower temperature. The corresponding entropy curves are shown in Fig. ~4. At $\Delta=2.3$, there develops a shoulder in the entropy function followed by a sharp downturn. NLC cannot capture the first order transition well, we expect the plateau-like behavior to be followed by a jump in entropy at the transition.

In Fig.~5, we show the uniform susceptibility as a function of temperature. For $\Delta=2.3$, there is a sudden rise in the susceptibility as the transition to high-spin states occurs. This sudden rise continues even in the RC phase as $\Delta$ is lowered further. This result is potentially relevant to experiments on BNOAS \cite{BNOAS-expt} as we discuss in the next section. The NLC results are not valid on the ordered side, the RC phase must have a finite susceptibility at $T=0$ \cite{CHN}.


\begin{figure}[htb!]
\centering
\includegraphics[width=\columnwidth]{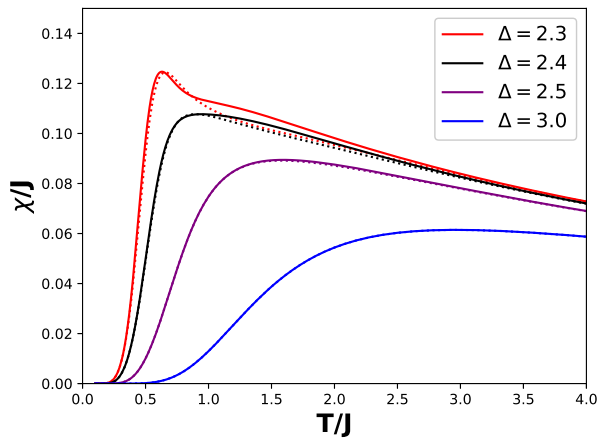}
\caption{Magnetic susceptibility as a function of temperature for various values of $\Delta$. Solid lines are 6th order NLC and dotted lines are 5th order NLC.}
\label{chi} 
\end{figure}

The above properties clearly distinguish our model from the Kondo-necklace, where a zero temperature quantum critical point turns into a quantum-critical fan at finite temperatures \cite{brenig, CHN}.

\section{V. Summary and Discussions}
In this paper we have studied the magnetic and thermodynamic properties of a two-electron two-orbital model system with competing high spin (S=1) and low spin (S=0) states, possibly relevant to the nickelates with a $d^8$ electronic configuration. The model has some similarities to the Kondo-necklace model proposed earlier by Jin et al \cite{jin}. However, detailed experimental properties are quite different and should be easy to distinguish experimentally. It is the purpose of the paper to bring out these differences. 

Being a two-dimensional model, there cannot be a Neel phase at finite temperatures. Instead, there is a Renormalized Classical phase for the antiferromagnet, which will develop Neel order at $T=0$, or at finite temperatures in presence of the slightest anisotropy or 3-dimensional coupling. This phase is separated by a first order phase transition from a non-magnetic spin-zero phase. The line of first order transitions terminates at a liquid-gas critical point. 

The magnetic behavior in the vicinity of the transition is quite unusual as the magnetic susceptibility can  be non-monotonic as a function of temperature and show a sudden rise as one transitions from a low-spin to high-spin phase. There can be hysteresis and meta-stability of the low and high spin phases. At the phase transition there should be a jump in entropy and a latent heat.

Jiang et al \cite{sawatzky} have discussed the importance of charge transfer energy and the tuning (and crossing) of singlet-triplet gap as a function of the charge transfer energy. Thus, it is possible that different materials, with small differences in ligand-environment, can be in different parts of the phase diagram. 
Some maybe magnetic, some non-magnetic and some may have parameters close to the critical value. 

The experimental results for the BNOAS material by Matsumoto et al \cite{BNOAS-expt} look very intriguing. There is a Neel transition at $130$ K. Just above the Neel transition at around $150$ K the system has a sudden increase in the magnetic susceptibility. The susceptibility increases by almost a factor of $6$ between $150$ and $130$ K. The author's speculate that this may be due to anisotropies that cause canted ferromagnetism. It is also possible that it is caused in part by the sudden emergence of high-spin from low-spin as seen in the present model. If true, this is direct evidence that the system is close to a degeneracy between high and low-spin states.

This manuscript is focused on the parent insulating materials. But, it is interesting to speculate about doping and superconductivity. From this point of view also this model may provide a sharp distinction to the proposal of Jin et al. Upon doping, the Kondo-necklace model should lead to the usual t-J model, whereas this model could lead to a spin-one or type II t-J model recently proposed by Zhang and Vishwanath \cite{ashvin}, with new mechanisms for superconductivity.


\section{Acknowledgments} 
We thank Warren Pickett for many useful discussions and for getting us interested in this problem. This work is supported in part by National Science Foundation grant number 
NSF-DMR 1855111.




\end{document}